\documentclass[a4paper,superscriptaddress,twocolumn,prd,showkeys,nofootinbib]{revtex4}
\newcommand{\figref}{Fig.~\ref}
\newcommand{\f}{\ensuremath{f_{\text{BBH}}^{\nu}}}
\newcommand{\fnull}{\ensuremath{f_0}}
\newcommand{\egw}{\text{3~M}\ensuremath{_{\odot}\ }}
\newcommand{\egwerr}{\ensuremath{3^{+0.5}_{-0.5}}\text{~M}\ensuremath{_{\odot}\ }}
\newcommand*\diff{\mathop{}\!\mathrm{d}}
\newcommand{\figwidth}{1}

\usepackage[caption=false]{subfig}

\usepackage{hyperref} 
\usepackage{graphicx,amsmath,lipsum}
\usepackage{siunitx}
\usepackage{longtable}
\usepackage{color}
 \usepackage[utf8]{inputenc}

\usepackage{tabularx} 
\newcolumntype{C}{>{\centering\arraybackslash}X} 

\begin{document}

\title{Constraints and prospects on GW and neutrino emissions using GW150914} 

\author{Krijn D. de Vries}
\affiliation{IIHE/ELEM, Vrije Universiteit Brussel, Pleinlaan 2, 1050 Brussels, Belgium}

\author{Gwenha\"el de Wasseige}
\affiliation{IIHE/ELEM, Vrije Universiteit Brussel, Pleinlaan 2, 1050 Brussels, Belgium}


\author{Jean-Marie Fr\`ere}
\affiliation{Universit\'e Libre Bruxelles, Boulevard de la Plaine 2, 1050 Brussels, Belgium}

\author{Matthias Vereecken}
\affiliation{IIHE/ELEM, Vrije Universiteit Brussel, Pleinlaan 2, 1050 Brussels, Belgium}
\affiliation{TENA, Vrije Universiteit Brussel, International Solvay institutes, Pleinlaan 2, 1050 Brussels, Belgium}


\begin{abstract}
Recently, the LIGO observatory reported the first direct observation of gravitational waves, with a signal consistent with a binary black hole merger. This detection triggered several follow-up searches for coincident emission in electromagnetic waves as well as neutrinos, but no such emission was found. In this article, the implications of the non-detection of counterpart neutrinos are investigated using general arguments. The results are interpreted with a parameter denoting the energy emitted in neutrinos relative to the energy emitted in gravitational waves. The bound on this parameter from the diffuse astrophysical neutrino flux detected by the IceCube Neutrino Observatory is discussed. It is found that, currently, the non-detection of counterpart neutrinos puts a bound comparable to the one from the diffuse astrophysical neutrino flux. This bound is then used to constrain the amount of matter in the black hole binary environment. Finally, the sensitivity to this parameter in future gravitational wave observation runs is investigated. It is shown how the detection of one or more neutrinos from a single merger would strongly constrain the source population and evolution. 
\end{abstract}

\maketitle

\section{Introduction and motivation}\label{intro}
On September 14th 2015, the two detectors of the Laser Interferometer Gravitational-Wave Observatory (LIGO) observed a transient gravitational wave signal, referred to as GW150914 \cite{Abbott:2016blz}. This signal matches with the expectations from the merger of two black holes with masses equal to 36$^{+5}_{-4}$ M$_ {\odot}$ and 29$^{+4} _{-4}$ M$_ {\odot}$.
The LIGO detection triggered a large follow-up campaign in both electromagnetic~\cite{Abbott:2016gcq} as well as neutrino~\cite{Adrian-Martinez:2016xgn, Abe:2016jwn, Aab:2016ras} detectors.
None of the follow-up searches triggered by GW150914 led to a significant detection\footnote{The Fermi-GBM has however reported a sub-threshold transient event, consistent with a short GRB \cite{Connaughton:2016umz}. It is still unclear whether this was associated with GW150914 or a chance coincidence.}. After the detection of GW150914 several models have been constructed that give rise to photon emission during a binary black hole (BBH) merger~\cite{Perna:2016jqh,Bartos:2016dgn,Kotera:2016dmp}. Nevertheless, in general no emission apart from gravitational waves (GW) is predicted, since no matter is expected to be present in the environment of the black hole binary. To test this hypothesis, given that Megaton-scale neutrino detectors such as IceCube~\cite{Aartsen:2016nxy}
, ANTARES~\cite{Spurio:2016sln} and Baikal-GVD~\cite{Avrorin:2016mva}
are available, it is useful to search for counterpart neutrinos. In view of the multi-messenger approach, a neutrino detection of a source discovered in gravitational waves would shine a unique light on the source properties.\\

In this paper, the potential of probing neutrino emission from GW sources from current and future GW events is investigated. A general approach is used which allows to constrain, for a given type of merger, the fraction of energy released in neutrinos relative to gravitational waves. The focus will be on the IceCube and ANTARES neutrino observatories and their energy range, above 100 GeV up to several PeV. 
At energies around the MeV-scale, neutrino emission could also be investigated. KamLAND~\cite{Gando:2016zhq} has published results of a search for MeV-neutrinos as counterpart of GW150914. However, it is likely that neutrinos at these energies would be produced by a completely different mechanism and matching the two results requires a detailed modelling of the $\nu$-spectrum.\\

In Section~\ref{section1}, the method is defined, followed by a brief discussion of the neutrino emission. It is shown how the diffuse astrophysical neutrino flux directly constrains the possible neutrino emission from BBH mergers. In Section~\ref{section2}, the method is applied to GW150914, with the focus on IceCube and ANTARES. Afterwards, the same method is repeated for multiple BBH mergers that could be detected in future observation runs. The gain in sensitivity and the reach by the end of LIGO run O2 is investigated. Finally, in Section~\ref{section3}, it is discussed how the current results are affected when considering different distributions of black hole masses. In addition, it is shown how the general results given here, can be interpreted using specific models. This immediately leads to a bound on the amount of matter in the black hole binary environment.
It should be stressed that, while the focus here is on BBH mergers, the parametrization is completely general and can also be used for other types of GW sources that might be discovered in the near future, such as neutron star-black hole and neutron star-neutron star mergers, for which neutrino emission is expected.

\section{Method}\label{section1}
\subsection{Neutrino emission strength}

The high-energy neutrino emission from the gravitationally detected BBH merger is investigated using the energy released through gravitational waves in combination with observations by neutrino telescopes. This will be applied to GW150914, which released \egwerr of energy into gravitational waves from a distance of 410$^{+160}_{-180}$~Mpc according to the LIGO analysis.\\

The amount of energy emitted in neutrinos within a certain energy range relative to the amount of energy emitted in gravitational waves (as reported by LIGO) is characterized by the neutrino emission fraction
\begin{equation}
\f = \frac{E_\nu}{E_{GW}}.
\label{eq:fractiondef}
\end{equation}
\\

There are different possibilities for the definition of \f. In the scenario presented above, it is implicitly assumed that there is an additional neutrino emission on top of the measured GW emission, with an energy of 
\begin{equation}
\begin{aligned}
& E_{\mathrm{GW}} = \egw, \\
& E_\nu = \f \times \egw,
\end{aligned}
\label{eq:case1}
\end{equation}
in the case of GW150914. Another scenario can be envisioned, where part of the energy loss of the binary is emitted in gravitational waves, while the other part is emitted in neutrinos and possibly other particles
\begin{equation}
\begin{aligned}
& E_{\mathrm{GW}} = \egw- \f \times \egw - X, \\
& E_\nu = \f \times \egw.
\end{aligned}
\end{equation}
In this case the definition of \f\ should be changed to contain the measured mass difference as reported by LIGO, instead of $E_{\mathrm{GW}}$. The most straightforward way to overcome this loss in signal strength is to perform a shift in the source distance. Note that the two cases imply different physics. The first case can represent neutrino emission coming from matter around the black holes, where \f\ can be larger than one. In the second case one considers more exotic scenarios, where part of the energy that would go to gravitational waves, is instead emitted in neutrinos. \\

If \f\ is small, the difference between the two cases becomes negligible. A large \f\ would imply a significant change in the physical conditions, either through added matter or by having weaker gravitational waves. In that case, the agreement between the measured signal and the general relativity simulations would likely be spoiled. Therefore, it will be assumed that \f\ is small. As will be shown in this paper, this assumption is valid.\\

High energy neutrino emission is typically associated with gamma-ray emission through pion decay. Whereas neutrinos can propagate unhindered, gamma-rays can be attenuated in a multitude of ways on their journey to Earth. To take this into account one needs to consider a specific model for the source environment. Therefore, in order to stay as general as possible, only the neutrino emission is treated, ignoring any constraints from gamma emission.

\subsection{Neutrino emission properties}

Two benchmark scenarios of neutrino emission are considered, both a mono-energetic spectrum as well as an $E^{-2}$-spectrum. The first scenario can be used when the neutrino spectrum is dominated by a single energy. It also allows for a direct convolution with any user defined spectrum. The second scenario is the standard power-law distribution that follows from Fermi acceleration. While there is no theoretical reason to expect a spectral index of exactly two for the situation considered in this work, it is the one corresponding to the high-energy flux given by IceCube~\cite{Aartsen:2015zva}. In both cases, the spectrum is normalized to $E_\nu$.\\

In the mono-energetic case, a scan is performed over the neutrino energy between \SI{100}{GeV} and \SI{100}{PeV} equal to the energy range of interest for IceCube and ANTARES. Since the number of neutrinos produced for mono-energetic production scales like $1/E$, while the interaction cross section for detection in this energy range increases with $E$~\cite{Formaggio:2013kya}, one expects that, up to detector effects, the amount of neutrinos detected at Earth is roughly constant. The same argument shows that, for general input spectra, the total number of detected neutrinos should be independent of the exact shape of the spectrum (assuming energy conservation). \\

When converting the emitted luminosity to the neutrino flux received at Earth, both the distance to the source (which is given by LIGO) and the angular distribution of the emission need to be considered. The luminosity distance given by LIGO has an associated uncertainty of about a factor of two. In the following, for simplicity, only the result of the central value is shown.\\

Gravitational waves from two merging black holes which are spiralling into each other, are emitted in all directions, with a slightly stronger flux along the angular momentum vector of the binary system. Therefore, the most likely orientation of a detected event is either face-on or face-off (see e.g.~\cite{TheLIGOScientific:2016pea}). In the case of jet-formation one expects the electromagnetic and neutrino emission to be beamed along this same direction. Therefore, it can be expected that, if there is emission other than gravitational waves, such emission would also be detectable. However, to stay general, all calculations will initially be done assuming isotropic emission. In the case of beaming, the flux will be enhanced with the beaming factor and the corresponding result can be directly obtained by rescaling from the isotropic case. An additional correction factor could also be included to take into account the possible different orientations of the source system. Redshift effects on the flux of individual events will be ignored in the following, which is reasonable in view of the current distance probed by LIGO. Finally, full mixing between the neutrino flavours is assumed, so that all three flavours arrive at Earth in equal amounts.

\subsection{Astrophysical bound}

In this section, the diffuse astrophysical neutrino flux first detected by IceCube in 2013~\cite{Aartsen:2013jdh} is used to put an upper bound on \f, following the approach in~\cite{Kowalski:2014zda,Ahlers:2014ioa}. Under the assumption that BBH mergers emit neutrinos throughout the history of the universe, the maximally allowed \f\ is the one which saturates the astrophysical neutrino flux. The rate of BBH mergers detectable by LIGO in the local universe is determined from all detected GW events so far\footnote{Besides GW150914, an additional binary black hole merger was detected, as well as a potential BBH merger in run O1~\cite{TheLIGOScientific:2016pea}.}. The 90\% credible interval is given by~\cite{TheLIGOScientific:2016pea}
\begin{equation}
R = \SIrange{9}{240}{Gpc^{-3} yr^{-1}}.
\label{eq:LIGOrate}
\end{equation}
To determine this range, in~\cite{TheLIGOScientific:2016pea} different black hole mass distributions are considered. In case of a mass distribution flat in log mass, given by $p(m_1,m_2)\propto\frac{1}{m_1m_2}$, the rate becomes,
\begin{equation}
R_{\mathrm{flat log}} = 31^{+42}_{-21} \mathrm{Gpc}^{-3} \mathrm{yr}^{-1}.
\label{eq:flatlograte}
\end{equation}
In case of a mass distribution following a power law equal to $p(m_1)\propto m_1^{-2.35}$, and $m_2$ uniform, the inferred rate becomes~\cite{TheLIGOScientific:2016pea},
\begin{equation}
R_{\mathrm{power law}} = 97^{+135}_{-67} \mathrm{Gpc}^{-3} \mathrm{yr}^{-1}.
\end{equation}
For both distributions, it was required that $5\ \mathrm{M}_\odot \leq m_2 \leq m_1$ and $m_1 + m_2 \leq 100\ \mathrm{M}_\odot$. 

For events with black hole masses similar to GW150914, the corresponding rate is given by~\cite{TheLIGOScientific:2016pea},
\begin{equation}
R_{\mathrm{GW150914}} = 3.4^{+8.8}_{-2.8} \mathrm{Gpc}^{-3} \mathrm{yr}^{-1}.
\label{eq:rate3m}
\end{equation}
\\

The diffuse neutrino flux resulting from a set of BBH mergers with properties similar to GW150914 will be considered. They produce gravitational waves with an energy of \egw along with an associated neutrino flux that follows an $E^{-2}$-spectrum between \SI{100}{GeV} and \SI{100}{PeV}. The corresponding rate of this class is given by $R$, for now unspecified. It will be discussed in Section~\ref{section3} how the results for a class of mergers with properties similar to GW150914 can be translated to results on the entire population of BBH mergers. The consequent diffuse neutrino flux is directly given by~\cite{Waxman:1998yy},
\begin{equation}
E^2 \left.\frac{\diff N_\nu}{\diff E_\nu}\right|_{\mathrm{obs}} = \left( \f t_H \frac{c}{4\pi} \xi_z\right) E^2 \left.\frac{\diff \dot{N}_\nu}{\diff E_\nu}\right|_{\mathrm{inj}, \f=1},
\label{eq:injtoflux}
\end{equation}
where
\begin{equation}
E^2 \left.\frac{\diff \dot{N}_\nu}{\diff E_\nu}\right|_{\mathrm{inj}, \f=1} = R\ E^2 \phi(E_\nu).
\label{eq:injflux}
\end{equation}
In here, $\phi(E_\nu)$ (in units of GeV$^{-1}$) is the total differential neutrino flux from an individual event with a total energy budget of $E_\nu = \egw$. The cosmic evolution of the sources is contained in $\xi_z$. 
Following~\cite{Ahlers:2014ioa}, one has
\begin{equation}
\xi_z(E) = \int_{0}^{\infty} \diff z\, \frac{H_0}{H(z)} \frac{\mathcal{L}_\nu(z, (1+z)E)}{\mathcal{L}_\nu(0, E)},
\end{equation}
where $H(z)$ is the redshift dependent Hubble parameter and $\mathcal{L}_\nu(z, E) = \mathcal{H}(z)\mathcal{Q}_\nu (E)$ is the spectral emission rate density. $\mathcal{H}(z)$ is the source density, with $\mathcal{H}(0)=R$, while $\mathcal{Q}_\nu (E)$ is the emission rate per source. For a power law ($\mathcal{L} \propto E^{-\gamma}$), $\xi_z$ is energy-independent.
For $\mathcal{H}(z)$ following the star formation rate (SFR)~\cite{Hopkins:2006bw,Yuksel:2008cu}, this results in $\xi_z \approx 2.4$. For no evolution in the local universe ($z<2$), it results in $\xi_z \approx 0.5$. In the following, $\xi=2.4$ will be used, unless stated otherwise. \\

The diffuse astrophysical neutrino flux measured by IceCube is given by~\cite{Aartsen:2015zva}
\begin{equation}
E^2 \Phi(E) = 0.84 \pm 0.3\times\SI{.e-8}{GeV\;cm^{-2} s^{-1} sr^{-1}},
\label{eq:icdiff}
\end{equation}
fitted with a fixed spectral index of 2 in the range between \SI{60}{TeV} and \SI{3}{PeV} of deposited energy. While there is a more up-to-date estimate of the flux~\cite{Aartsen:2016xlq}, which was fitted with a free spectral index, the analysis is performed using the standard spectral index of 2. Since the assumed neutrino spectrum is valid over an energy range wider than the one where this spectrum was measured, the upper bound on \f\ is found when the fluxes predicted by Eq.~\eqref{eq:injtoflux} and observed by IceCube in Eq.~\eqref{eq:icdiff} are equal, i.e. when the normalization constants are equal.\\

The resulting bounds on \f\ will be calculated for two source classes. When considering only the neutrinos emitted by BBH mergers similar to GW150914, the rate given in Eq.~\ref{eq:rate3m} is used to give
\begin{equation}
\f \lesssim 3.63_{-2.62}^{+17.0}\times10^{-3}.
\label{eq:astrobound_3m}
\end{equation}
It is also possible to consider the full mass distribution of BBH mergers. Since this distribution is not known, there is instead a range of merger rates (Eq.~\ref{eq:LIGOrate}), resulting in a bound on \f\ between
\begin{equation}
\f \lesssim \numrange{5.15e-5}{1.37e-3}.
\label{eq:astrobound_LIGO}
\end{equation}
The first of these bounds (Eq.~\ref{eq:astrobound_3m}) will be used when the bounds from GW150914 itself are investigated in Section~\ref{sec:GW150914limits}, while the second (Eq.~\ref{eq:astrobound_LIGO}) will be used to compare with the prospective bound from a population of detected BBH mergers in Section~\ref{sec:prospects}.

It should be noted that as further GW events are detected by LIGO and Virgo, the BBH mass distribution and typical $E_{\mathrm{GW}}$ will be known with more precision. This will allow the present bound to be calculated more accurately.\\

If BBH mergers emit neutrinos with a mono-energetic spectrum, the results change. The diffuse neutrino spectrum from these BBH mergers will follow the redshift evolution of the source, instead of a simple power-law spectrum. We therefore restrict ourselves to an $E^{-2}$ emission scenario for the astrophysical bound.

\section{Detection of GW neutrinos}\label{section2}

\subsection{Limits from GW150914}\label{sec:GW150914limits}

In order to show how the non-detection of counterpart neutrinos\footnote{Three neutrino events were detected in the 1000~s time window around GW150914, which were outside the 99\% confidence region given by LIGO. This number being compatible with the background expectation, we consider that no signal events were detected.} from GW150914 constrains the neutrino emission fraction \f, it is necessary to convert the flux at Earth to the flux seen in a detector. The present analysis will focus on both IceCube~\cite{Aartsen:2016nxy} 
and ANTARES~\cite{Spurio:2016sln}, which can detect high energy neutrinos between \SI{100}{GeV} and \SI{100}{PeV}. In order to be similar to the follow-up search of GW150914 by IceCube~\cite{Adrian-Martinez:2016xgn}, the IceCube effective area presented in~\cite{Aartsen:2014cva} will be used. Such an effective area is obtained from a search for muon neutrinos, because of their excellent pointing. Assuming full mixing between the neutrino species, this means that the flux of interest is roughly a factor 3 smaller. The IceCube effective area is given for three declination bands in the Southern Sky ($-90^\circ < \delta < -60^\circ$, $-60^\circ < \delta < -30^\circ$ and $-30^\circ < \delta < 0^\circ$). The IceCube analysis is such that the background rate is uniform over the entire sky. In a time window of \SI{1000}{s} around GW150914, which can be assumed to contain the full neutrino signal, the expected background is 2.2 events over the full Southern Sky~\cite{Adrian-Martinez:2016xgn}. Similarly, the ANTARES effective area presented in ~\cite{AdrianMartinez:2011uh} will be used, which is given for two declination bands in the Southern Sky ($-90^\circ<\delta<-45^\circ$ and $-45^\circ < \delta < 0^\circ$). From this, ANTARES expected to see 0.014 neutrino events in the Southern Sky in a time window of \SI{1000}{s} around GW150914~\cite{Adrian-Martinez:2016xgn}. The localization of GW150914 is such that it is spread out over the Southern Sky.\\

\begin{figure}
	\centering
	\includegraphics[width=\figwidth\linewidth]{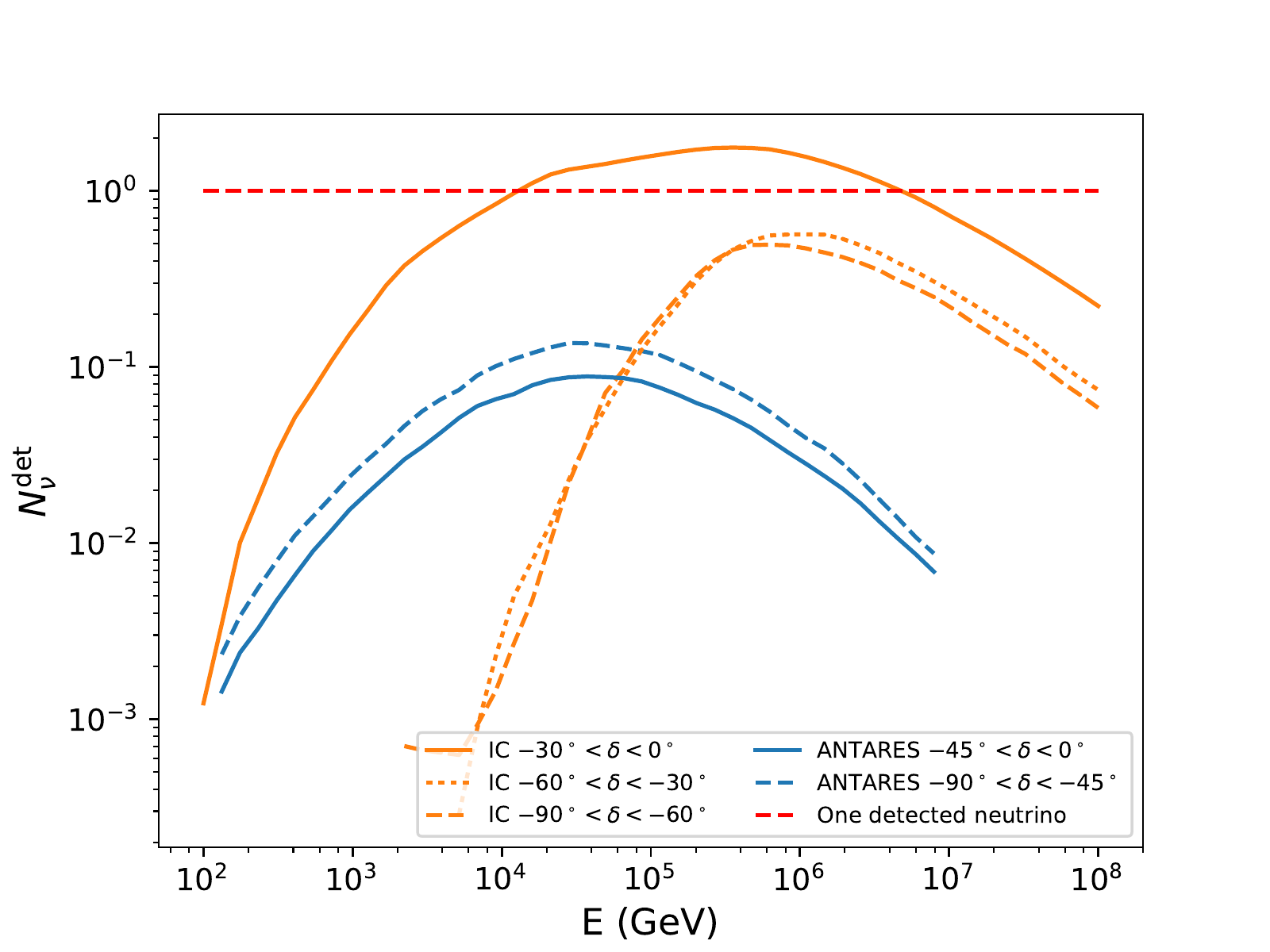}
	\caption{The amount of neutrinos detectable for a single GW event similar to GW150914 in the case of isotropic mono-energetic emission for $\f=\num{.e-2}$ as a function of neutrino energy. The results are shown for both IceCube and ANTARES and for different declination bands.
	}
	\label{fig:monodetected}
\end{figure}

In \figref{fig:monodetected}, the number of detectable neutrinos is shown for a BBH merger similar to GW150914, located in the Southern Sky given a neutrino energy fraction equal to $\f=\num{.e-2}$. The obtained values are given for isotropic, mono-energetic emission between \SI{100}{GeV} and \SI{100}{PeV}. It follows that the IceCube sensitivity drops towards the more southern declination bands, which can be understood by the atmospheric muon background which becomes increasingly larger for this part of the sky. ANTARES on the other hand, since it is located in the Northern Hemisphere, is shielded for atmospheric muons for this part of the sky and only has to cope with the atmospheric neutrino background. As such, for the most southern part of the sky, at energies below ~10 TeV, the ANTARES sensitivity becomes leading.\\

From the different results in \figref{fig:monodetected}, it can be seen that, for monoenergetic neutrino emission and for constant \f, the number of detectable neutrinos varies little between \SI{.e4}{GeV} and about \SI{.e7}{GeV}. Outside this range, the sensitivity is affected by detector energy resolution in the lower end and limited statistics in the upper end of the energy range. 
Indicated on the figure is the single neutrino detection threshold (dashed red line). It follows that the non-detection of counterpart neutrinos for GW150914 puts a bound equal to $$\f\lesssim\num{.e-2},$$ in an energy range between \SI{.e4}{GeV} and \SI{.e7}{GeV} for mono-energetic emission when considering the effective area near the horizon. For the more southern effective area, the bound is weakened. \\


Next, the constraints for the more standard case of an $E^{-2}$-spectrum will be investigated. Given that LIGO localizes GW150914 in an area spread out over the entire Southern Sky and considering the difference in effective area for the various declination bands, the analysis will be done for two extreme cases. Firstly, for the declination band $-30^\circ < \delta < 0^\circ$, the effective area of IceCube is the largest since in this region the atmospheric muon background is still relatively small. The resulting bound for this region is optimistic and will be used test whether a neutrino signal could have been seen even in the best case scenario for viable models of neutrino emission from BBH mergers. Secondly, in order to have a conservative bound on \f, the effective area in the declination band $-90^\circ < \delta < -60^\circ$ will also be considered. Here, ANTARES has a larger effective area in the low energy range, while the one of IceCube is larger in the high energy range. Since a combined analysis is beyond the scope of this work, the energy range is instead split in two regimes, so that in each energy regime the experiment with the largest effective area is used.\\

For the calculation of the astrophysical bounds, only the sub-class of binary black hole mergers that is similar to GW150914 will be considered. All of these mergers emit \egw of energy in gravitational waves, with a rate as in Eq.~\ref{eq:rate3m}. The resulting bound on \f\ was already given in Eq.~\ref{eq:astrobound_3m}. Since the contribution from BBH mergers with different properties are not taken into account, this leads to a conservative bound on \f.\\

\begin{figure*}
	\centering
	\subfloat[Results using only IceCube, for the effective area in the declination band $-30^\circ < \delta < 0^\circ$. This is the most sensitive region and leads to the most optimistic bound on \f.]{\includegraphics[width=.49\linewidth]{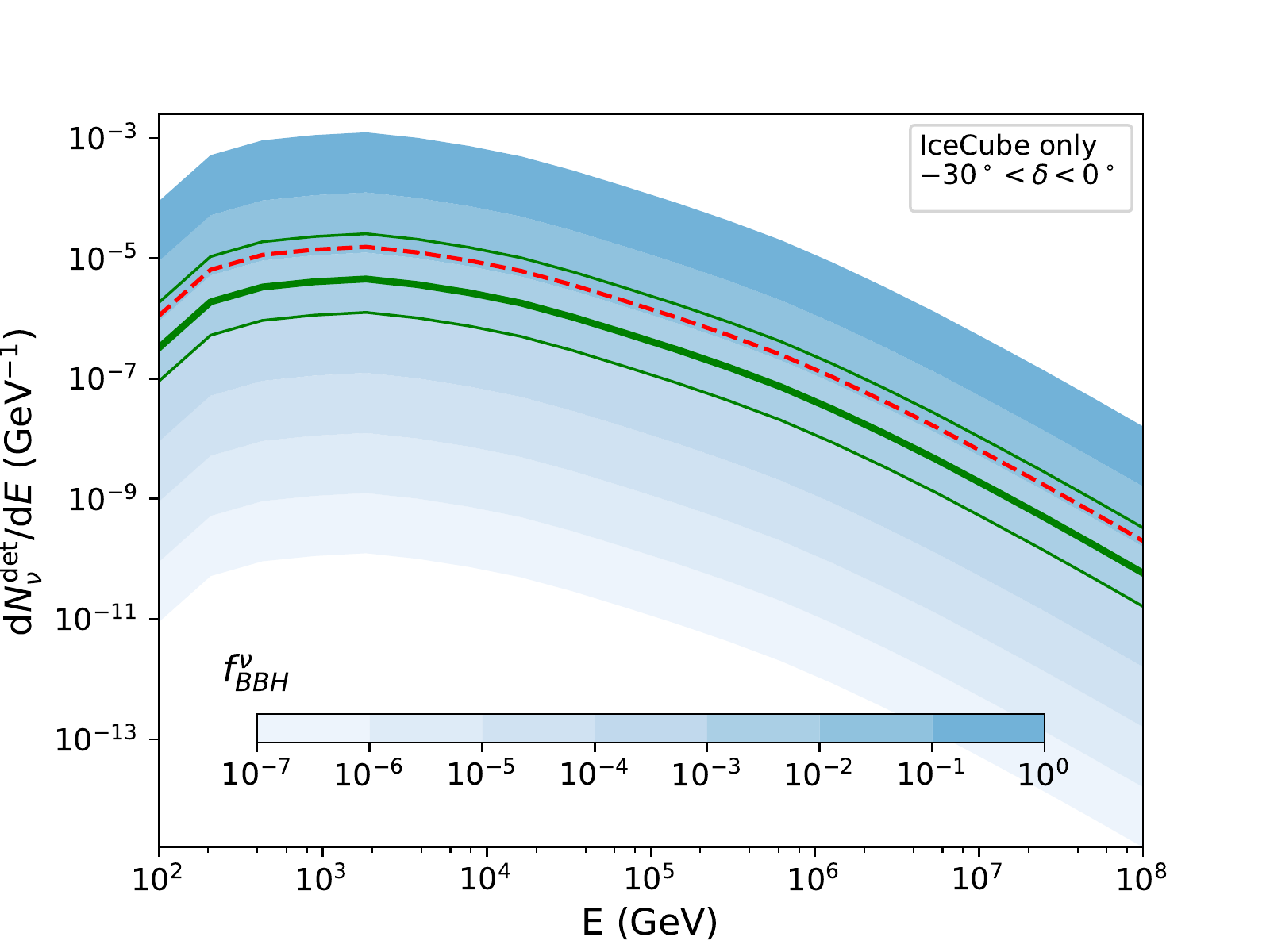}}	
	\hfill
	\subfloat[Results using the effective area in the declination band $-90^\circ < \delta < -60^\circ$, which is the least sensitive region. ANTARES and IceCube effective areas are used in the energy range where the respective experiment is the more sensitive.]{\includegraphics[width=.49\linewidth]{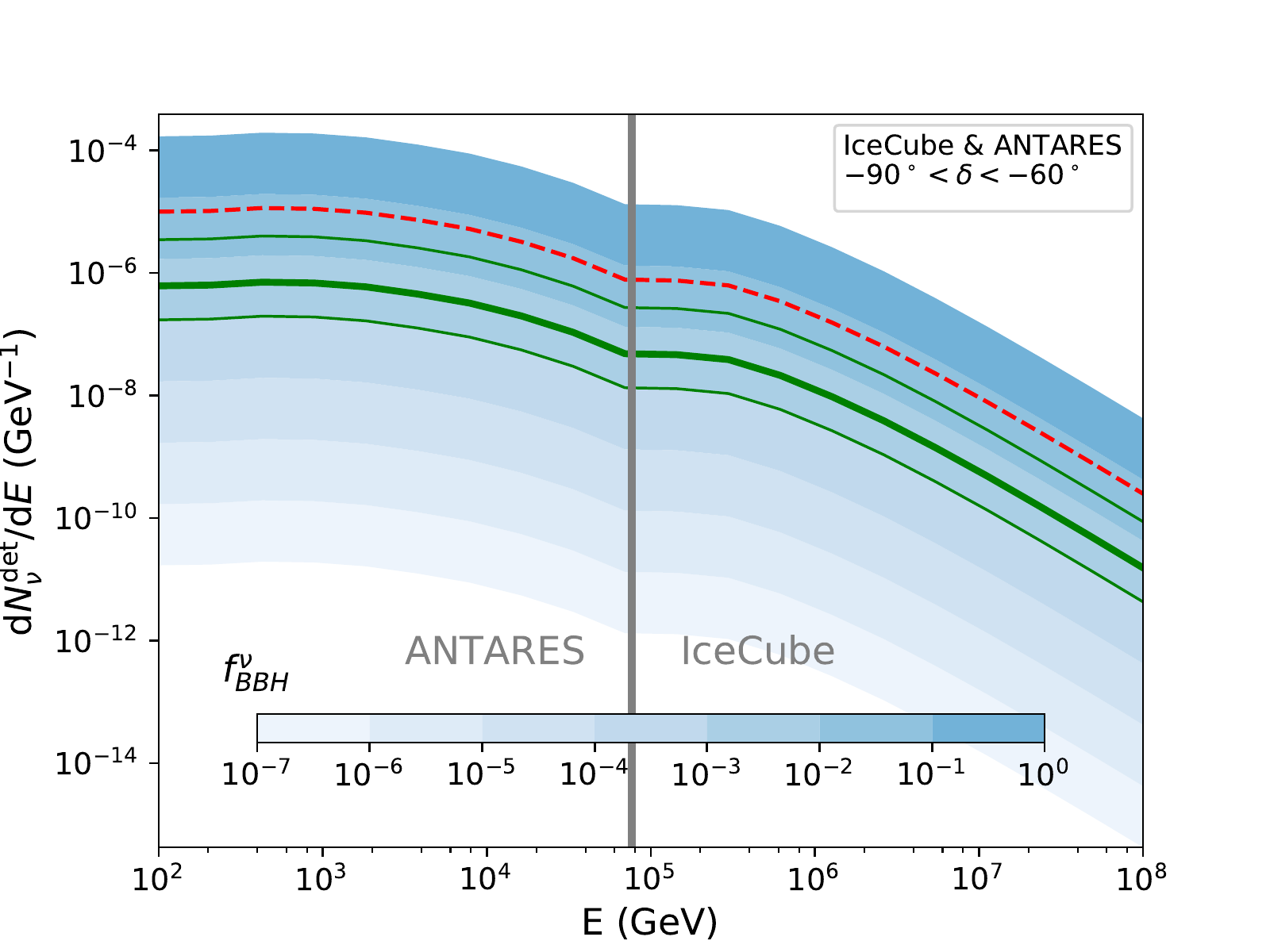}}
	\caption{The detected flux for a single GW event similar to GW150914 in the case of an isotropic $E^{-2}$-spectrum, for different \f\ (blue band).
The red dashed line shows the flux for which one event is detectable for this BBH merger event. The green lines show the upper bound from the astrophysical neutrino flux and its uncertainty for the class of BBH mergers similar to GW150914 (Eq.~\ref{eq:rate3m})}.
	\label{fig:detectedE2fluxisotropic}
\end{figure*}

In~\figref{fig:detectedE2fluxisotropic}, the predicted flux assuming an isotropically emitted $E^{-2}$ spectrum is shown for different values of the neutrino energy fraction \f\ ranging from~\num{.e-7} to \num{1}, indicated by the blue bands. The red dashed line again indicates the threshold where a single event detection would be detected integrated over the entire energy range. It follows that the non-detection of a neutrino counterpart from GW150914 puts an optimistic bound $$\f\lesssim \num{1.24e-02},$$ using the effective area in the declination band $-30^\circ < \delta < 0^\circ$, and a conservative bound $$\f\lesssim\num{5.89e-02},$$ using the effective area in the declination band $-90^\circ < \delta < -60^\circ$. The resulting bounds for the different cases therefore show little difference. 
As previously stated, the astrophysical bound has a value of $\f \lesssim 3.63_{-2.62}^{+17.0}\times10^{-3}$ (Eq.~\ref{eq:astrobound_3m}) and therefore stands below the single event detection threshold. \\

\begin{figure}
	\centering
	\includegraphics[width=\figwidth\linewidth]{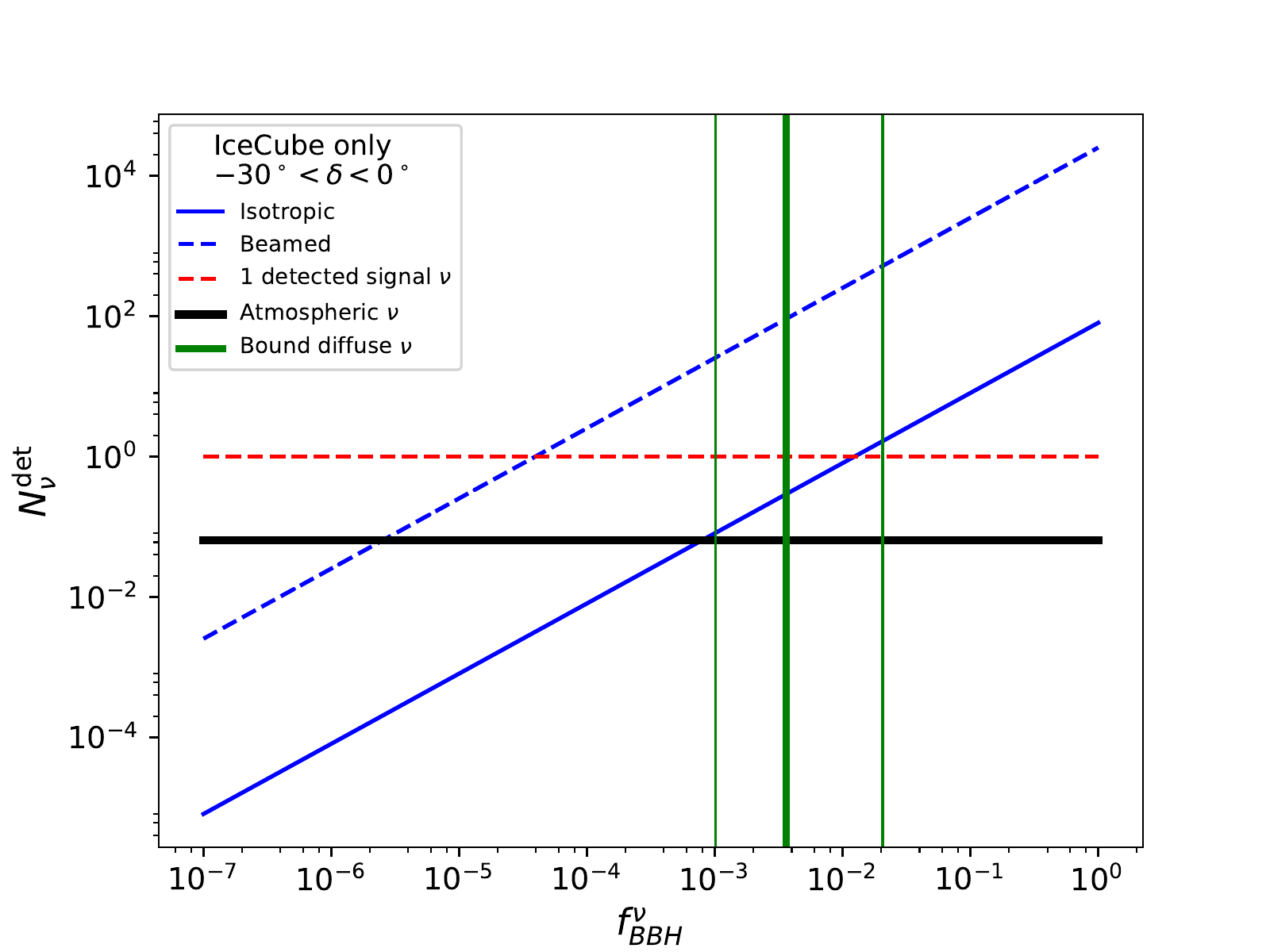}
	\caption{The integrated number of events from a source with the properties of GW150914, following an $E^{-2}$-spectrum for different \f, with isotropic emission as a full blue line and beamed emission as the dashed blue line. The fat black line is the atmospheric neutrino flux, which is integrated over a time window of \SI{1000}{s} and a solid angle of 600~deg$^2$. The red dashed line shows the one detectable event threshold. The green lines show the upper bound from the astrophysical neutrino flux and its uncertainty for the class of BBH mergers similar to GW150914 (Eq.~\ref{eq:rate3m}).}
	\label{fig:integratedE2events}
\end{figure}

\figref{fig:integratedE2events} shows the integrated number of events one expects from a source class with the properties of GW150914 as a function of the neutrino energy fraction \f, for both isotropic (full blue line) and beamed (dashed blue line) emission. To investigate which \f\ could have lead to a visible neutrino signal in the most optimistic case, the effective area of IceCube near the horizon will be used. The expected number of background events is 2.2 in a time window of \SI{1000}{s} around GW150914 for the entire Southern Sky~\cite{Aartsen:2014cva}. This number can then be rescaled to a solid angle of 600~deg$^2$, which corresponds to the localization of GW150914. The resulting background, which is shown by the full black line, is negligible for a single event.
The bound from non-detection can be read off from the crossing of these blue lines with the one detected event threshold given by the red dashed line. In the case of beaming, the flux towards Earth can be enhanced. For example, if a jet emits in a patch of $\Delta\Omega=0.2 \times 0.2$ in solid angle, the flux would be enhanced with a factor $\frac{4\pi}{0.2\times0.2}$. As can be read off from Fig.~\ref{fig:integratedE2events}, the one detectable event threshold in the case of a beamed $E^{-2}$-spectrum would then become $${\f=\num{3.96e-5} \times \frac{\Delta\Omega}{0.2\times0.2}}.$$ The astrophysical flux would not change when individual sources have a beamed emission, since the diffuse flux is still isotropic. 
It follows that the limits on \f\ obtained from the non-detection of counterpart neutrinos from GW150914 are stronger than those obtained from the astrophysical flux assuming a beamed emission close to the equator. This immediately implies that in case BBH mergers similar to GW150914 are responsible for the astrophysical neutrino flux, that either the emission from GW150914 was not beamed towards Earth, or that the beaming was smaller than $\Delta\Omega = 3.68~\mathrm{sr}$.\\

Since no neutrino has been detected so-far and the single event detection threshold for isotropic emission is above the astrophysical bound, currently all source populations are still allowed. In case of a neutrino detection in the near future, \f\ will still be above the given astrophysical bound. This would imply that the assumptions that went into this bound are too strong, so that merger rate, injected energy and source evolution are constrained. This point will be elaborated upon in the following sections.\\

\subsection{Prospects}
\label{sec:prospects}


In LIGO run O2, it is expected that more BBH mergers will be seen. Here, it is investigated how a stacked search of these events can constrain \f, assuming the more realistic $E^{-2}$ neutrino emission spectrum. Since the full BBH population is considered, the rate estimate in Eq.~\ref{eq:LIGOrate} will be used. As a result both a high and a low astrophysical bound will be shown, corresponding to Eq.~\ref{eq:astrobound_LIGO}. Furthermore, it is assumed all the BBH mergers will be similar to GW150914, radiating \egw in gravitational waves from a distance of 410~Mpc. These assumptions will be discussed in Section~\ref{section3}. Since GW150914 is expected to be among the more powerful BBH mergers that could occur and it is relatively close by, this leads to an estimate of the smallest \f that can potentially be probed.\\

The details of the analysis are similar to the one in the previous section, with minor adjustments. Because the mergers could happen anywhere in the sky, the IceCube effective area is averaged over the full sky. The localization of GW events is expected to improve with the improvements and enlargement of the LIGO-Virgo network~\cite{Gehrels:2015uga}.
%
%
%
%
%
%
%
%
%
In that case, neutrino observatories will be able to limit their search to a smaller solid angle in the sky, resulting in a reduced background. Therefore, the calculation is done for a localization of 600~deg$^2$, 100~deg$^2$ and 20~deg$^2$.
Only the irreducible background from atmospheric neutrinos~\cite{Sinegovskaya:2013wgm} is considered. This simplification that corresponds to the case of an ideal analysis, is also representative for the near future situation where KM3NeT~\cite{Adrian-Martinez:2016fdl} and Baikal-GVD~\cite{Avrorin:2016mva} will be online, and both the Northern and Southern Sky will be optimally observed.
The background is integrated over \SI{1000}{s}, which is the time window considered in the GW follow-up analysis. This conservative time window allows for the assumption that the full neutrino signal is contained.\\

\begin{figure}
	\centering
	\includegraphics[width=\figwidth\linewidth]{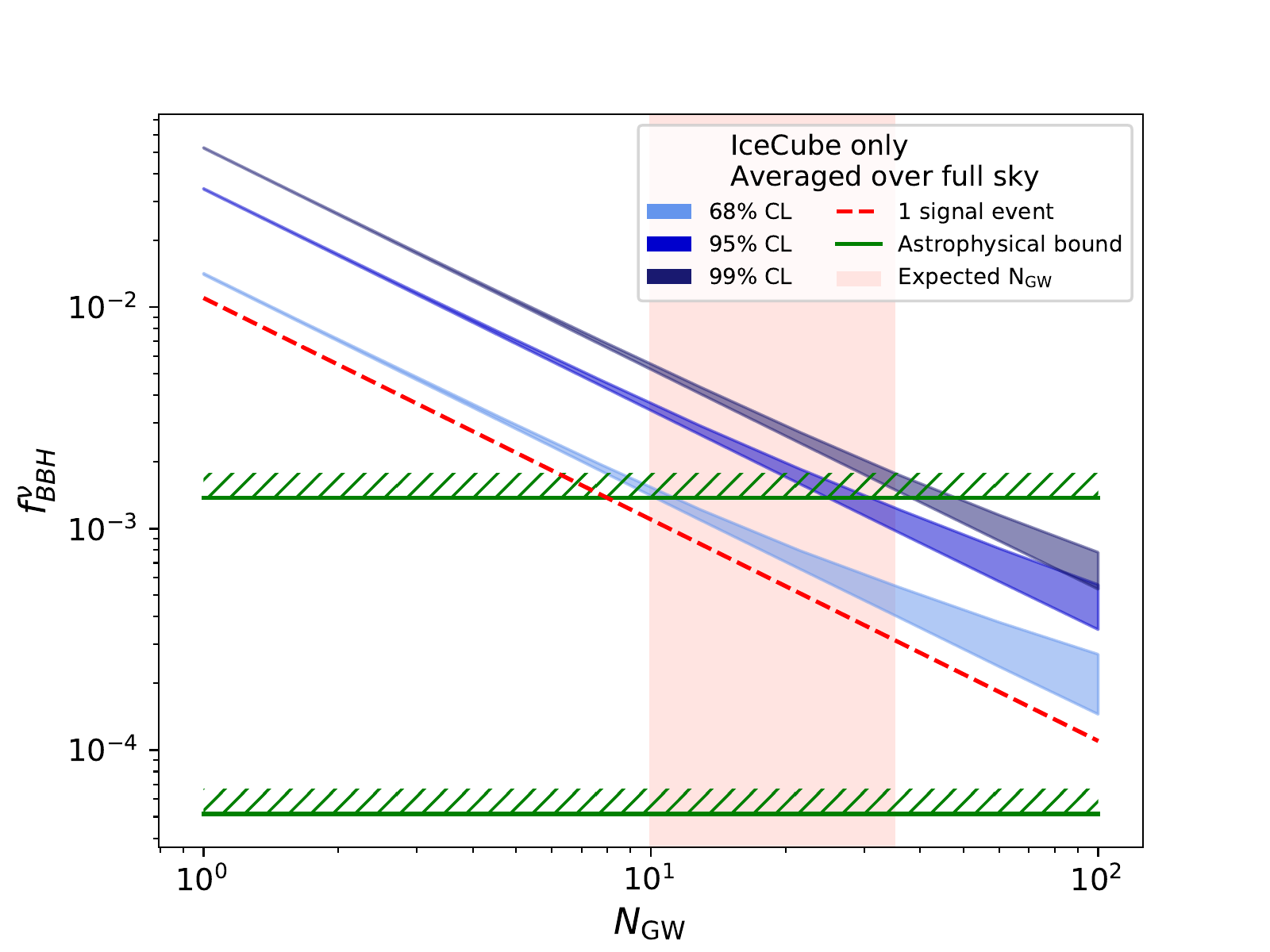}
	\caption{The expected average upper limits on \f\ at 68\%, 95\% and 99\% CL (from bottom to top) as a function of number of \egw BBH mergers observed in gravitational waves, using the IceCube effective area averaged over the full sky. Depending on the LIGO accuracy in locating the event, the IceCube background rejection varies and results in upper limit bands for localizations between 600~deg$^2$ and 20~deg$^2$. The green hatched lines show the upper bounds from the astrophysical neutrino flux for the upper and lower limit of the BBH merger rate for the full population of BBH mergers (Eq.~\ref{eq:LIGOrate}). The vertical band shows the expected number of BBH mergers seen in LIGO run O2.}
	\label{fig:fracpotentialAVUL}
\end{figure}

\figref{fig:fracpotentialAVUL} shows the average upper limits on \f\ at 68\%, 95\% and 99\% confidence level (blue bands) that can be expected as a function of the number of detected BBH merger events ($N_{\mathrm{GW}}$) by LIGO. The calculation of the upper limits follows the approach in~\cite{Hill:2002nv}. The bands indicate the possible improvement of the localization BBH merger events from 600~deg$^2$ to 20~deg$^2$.
The red dashed line indicates at what \f\ at least one signal neutrino can be detected, integrated over all BBH merger events. At first, the limit on \f\ drops proportionally to the single event detection threshold, since the detection is purely signal limited. Starting at around 10 BBH mergers, however, the background starts to become significant and the limit drops less fast.
It is at this point that the improved localization starts to become important. \\

The obtained values for \f\ can be compared with the astrophysical bounds corresponding to the upper and lower limits of the BBH merger rates given in~Eq.~\ref{eq:LIGOrate}, which, following Eq.~\ref{eq:astrobound_LIGO}, are equal to $\f \lesssim \num{1.37e-03}$ and $\f \lesssim \num{5.15e-05}$, shown by the hatched green lines. It should be noted that as more BBH mergers are observed, the estimate of the rate will improve, so that these two astrophysical bounds should get closer. It is found that the average upper limits on \f\ reach the highest astrophysical bound at 
$$N_{\mathrm{GW}}\gtrsim 10,\,12,\,14,$$ at 68\%, 95\% and 99\% CL
respectively, with small differences between the different uncertainties in the localization. If signal neutrinos would be found before reaching this number of BBH mergers, the source population (merger rate and cosmic evolution of the sources) would be strongly constrained by the diffuse astrophysical neutrino flux. The average upper limit from a search for counterpart neutrinos only reaches the lowest astrophysical bound for
$$N_{\mathrm{GW}}\gtrsim 300,$$ at 68\% CL
and for a localization of 20~deg$^2$.
The vertical band indicates the expected number of BBH merger observations at the end of LIGO run O2, which is between \numrange{10}{35}~\cite{Abbott:2016nhf}. A wider estimate puts this number between \numrange{2}{100}, which covers the whole plot. It follows that the number of GW events needed to constrain the lowest astrophysical bound is well outside the reach of LIGO run O2. By the end of run O2, if 
10 BBH mergers would be observed, it would be possible to limit \f\ down to about $$\f\approx  \num{1e-3},\,\num{4e-3},\,\num{6e-3},$$ at 68\%, 95\% and 99\% CL respectively.
If indeed 35 BBH mergers would be observed, it would be possible to limit \f\ down to about 
$$\f\approx  \num{5e-4},\,\num{1e-3},\,\num{2e-3},$$ at 68\%, 95\% and 99\% CL
respectively. \\


\begin{figure}
	\centering
	\includegraphics[width=\figwidth\linewidth]{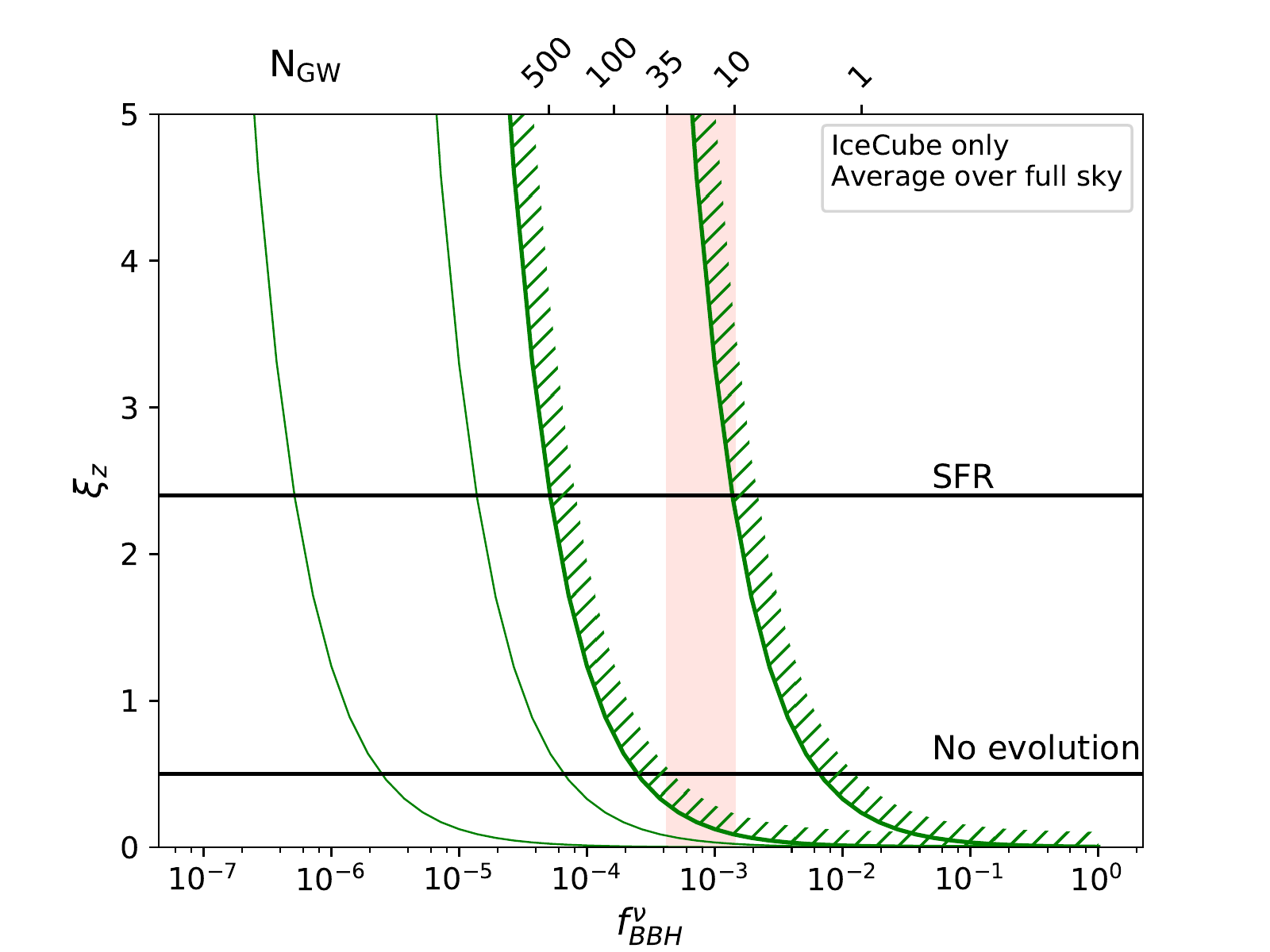}
	\caption{The $\xi_z$-\f\ plane that can be constrained with the combination of the astrophysical diffuse flux (hatched green lines) and direct searches for coincident $\nu$'s. The horizontal axis on top shows the number of gravitational wave events similar to GW150914 ($N_{\mathrm{GW}}$) necessary to see the corresponding value of \f\ at 
68\% CL (as in Fig.~\ref{fig:fracpotentialAVUL} and with a 100~deg$^2$ resolution). Also shown are the bounds when BBH mergers can only be responsible for 1\% of the diffuse astrophysical neutrino flux (thin green lines). The vertical band shows the expected number of BBH mergers seen in LIGO run O2. The two values of $\xi_z$ corresponding to a source evolution following the star formation rate and no evolution are indicated by black lines.}
	\label{fig:xizplot}
\end{figure}

Following~Eq.~\ref{eq:injtoflux}, there is a degeneracy between the neutrino energy fraction \f\, and the source evolution parameter $\xi_z$. To illustrate this degeneracy, in~Fig.~\ref{fig:xizplot}, the $\xi_z$-\f\ plane is shown. The constraints from the direct neutrino searches are given on the top-axis. Hence, the current constraint from the non-detection of a neutrino counterpart from GW150914 is given by $N_{GW}=1$ and the possible constraints after LIGO run O2 are indicated by the red band. The solid green lines indicate the bounds where the GW neutrino flux would saturate the astrophysical neutrino flux detected by IceCube. It follows that if a single counterpart neutrino event would have been observed, or is observed within 10 GW events, the astrophysical flux can only be explained for source evolutions $\xi_z < 3$. Given the current uncertainties on the BBH merger rate, to rule out BBH mergers as the main sources for the astrophysical neutrino flux, one needs to detect at least 1000 BBH mergers. Nevertheless, assuming that the BBH merger rate is determined accurately to its central value after several detections, this might already be achieved after the 10-35 events predicted for LIGO run O2. To illustrate the level at which BBH mergers can be excluded as the source for the diffuse high-energy astrophysical neutrino flux, in~Fig.~\ref{fig:xizplot}, the bounds (thin green lines) where the BBH merger neutrino flux would correspond to 1\% of this flux are also indicated.


\section{Discussion}\label{section3}

\subsection{Population of BBH mergers}

The arguments presented in this paper should be robust, general and lead to order of magnitude estimates for the neutrino emission fraction \f. 
In order to extend the predictions on \f\ to BBH mergers of varying black hole masses (and thus varying $E_{\mathrm{GW}}$), one has to make an assumption on the scaling of the neutrino emission for these different masses. The simplest assumption is that $E_\nu \propto E_{\mathrm{GW}}$, so that \f\ is a \textit{universal} fraction for all binary black hole mergers.
This assumption is valid, for example, if both $E_{\mathrm{GW}}$ and $E_\nu$ are proportional to the sum of the masses of the black holes. This is a reasonable approximation for $E_{\mathrm{GW}}$, since it is true for equal-mass non-spinning black holes in the inspiral phase, as the released energy is proportional to the reduced mass of the binary. The validity of this approximation was checked using fits to numerical simulations of non-spinning binary black hole mergers~\cite{Berti:2007fi,Baker:2008mj}.
For $E_\nu$, this scaling depends on the origin of the neutrino emission. In the case of a GRB-like scenario, the matter that seeds the neutrino production is a remnant of the original star that formed the black hole. There, the assumption that the amount of matter available scales linearly with the star (and black hole) mass is reasonable. \\

It is also possible to consider more general relations between $E_{\mathrm{GW}}$ and $E_\nu$.
This is illustrated by decomposing \f,
\begin{equation}
\f=\fnull\times g\left(\frac{\mathrm{M}_\mathrm{BH}}{\mathrm{M}_{\mathrm{GW150914}}}\right).
\end{equation}
Here the normalized mass function $g\left(\frac{\mathrm{M}_\mathrm{BH}}{\mathrm{M}_{\mathrm{GW150914}}}\right)$ includes the amount of matter available to produce neutrinos, where in this article only a dependence on the combined mass of the black holes $M_{\mathrm{BH}} = m_1 + m_2$ is considered.\\

To illustrate the effect of such a scaling, one can consider the situation described in the previous sections. To obtain the diffuse neutrino flux, the emission per source has to be convoluted with the black hole mass distribution. By considering a black hole mass distribution flat in log mass ($p(m_1,m_2) \propto \frac{1}{m_1m_2}$), in combination with a neutrino emission proportional to $M_{\mathrm{BH}}$ (i.e. $g=1$ and \f\ universal), the diffuse neutrino emission is roughly independent of the black hole masses. Therefore, the results presented in Fig.~\ref{fig:fracpotentialAVUL}, assuming all BBH mergers would be similar to GW150914, resemble the realistic situation of a flat in log mass BBH distribution, in combination with a neutrino emission that scales linearly with the black hole mass. The energy fraction is now calibrated by $\f(M_{\mathrm{BH}}=M_{\mathrm{GW150914}})=\fnull$.\\

Another possible situation could be inverse mass scaling such that $E_\nu \propto 1/M_{\mathrm{BH}}$, which leads to
\begin{equation}
\f\left(\mathrm{M}_{\mathrm{BH}}\right) = \fnull \times \left(\frac{\mathrm{M}_\mathrm{GW150914}}{\mathrm{M}_{\mathrm{BH}}}\right)^2.
\end{equation}
In case of a flat in log mass distribution, the neutrino emission from high-mass black hole binaries will be further suppressed, hence the neutrino emission in this situation will be dominated by low-mass black hole mergers.
\\

It follows that if one is able to determine the neutrino energy fraction \f\ for different sub-classes of BBH mergers, the internal neutrino emission properties, as well as the source environment are directly probed. An example of a specific sub-class is given in~\cite{Bartos:2016dgn}. Here BBH mergers in active galactic nuclei are considered, where it is shown that one might expect an enhanced neutrino energy fraction \f\ for this source class.

\subsection{Model-dependent interpretation}

The results presented here can be used to draw more model specific conclusions on the neutrino production. In general, no neutrino emission is expected from BBH mergers, since the black holes should have cleared the environment of all matter long before the merger occurs. This statement can be tested, by assuming neutrinos are produced by accelerated matter around the BBH, corresponding to the case of Eq.~\ref{eq:case1}. One can then decompose \f\ as
\begin{equation}
\f = f_{\mathrm{matter}}\times f_{\mathrm{engine}} \times \epsilon_{\mathrm{p, acc}} \times \epsilon_{\nu}.
\label{eq:fparts}
\end{equation}
Herein $f_{\mathrm{matter}}$ denotes the amount of matter present around the BBH relative to the amount of energy emitted in gravitational waves. The acceleration model is contained in the combination $f_{\mathrm{engine}} \times \epsilon_{\mathrm{p, acc}}$. The first of these, $f_{\mathrm{engine}}$, contains the amount of energy which is put into an acceleration engine, relative to the amount of matter present. The second of these, $\epsilon_{\mathrm{p, acc}}$, reflects the amount of protons which can be accelerated to high energy.  The fraction of energy from the accelerated particles that ends up in neutrinos in the considered energy range is given by $\epsilon_\nu$. \\

As an example, assume that the situation of an accretion disk falling into a BBH is similar to matter from a neutron star falling onto a companion black hole. In this situation, the neutrino production mechanism is similar to the GRB-fireball model~\cite{Waxman:1997ti}. The conversion factor from accretion disk mass to fireball energy (primarily from potential energy) is expected to be of order $f_{\mathrm{engine}}=1/10$. The amount of energy from the fireball that goes into the protons is given by $\epsilon_{\mathrm{p, acc}}=1/10$. Finally, the amount of energy in protons that goes into neutrinos is given by $\epsilon_\nu=1/20$. Hence 
\begin{equation}
\f \lesssim  f_{\mathrm{matter}}\times 5\cdot10^{-4}.
\end{equation}
This allows then to immediately constrain the amount of matter surrounding the two black holes. Using the non-detection of counterpart neutrinos in GW150914, which was in the most optimistic case at $\f= \num{3.96e-05}$ for beamed emission in a typical solid angle $\Delta \Omega=0.2\times0.2$ directed towards Earth, this results in
\begin{equation}
f_{\mathrm{matter}}^{GW150914} \lesssim \num{7.9e-2} \times \frac{\Delta \Omega}{0.2\times0.2}.
\end{equation}
In addition, the expected limit after 10 and 35 BBH mergers detected by LIGO can also be used. From the analysis in Section~\ref{sec:prospects}, the expected limit on the amount of matter in the black hole binary environment is
\begin{equation}
f_{\mathrm{matter}}^{N_\mathrm{GW}=10} \lesssim \num{6e-3} \times \frac{\Delta \Omega}{0.2\times0.2}.
\end{equation}
\begin{equation}
f_{\mathrm{matter}}^{N_\mathrm{GW}=35} \lesssim \num{3e-3} \times \frac{\Delta \Omega}{0.2\times0.2}.
\end{equation}
From Fig.~\ref{fig:fracpotentialAVUL}, it follows that the astrophysical limits are weaker than the limits obtained from the non-detection of counterpart neutrinos from GW150914 in case of a beamed emission. As such, only the latter is considered for the limit on $f_\mathrm{matter}$.
\\

To get an estimate of \f\ for possible neutrino emission coming from BBH mergers using the GRB-fireball mechanism, consider the model in~\cite{Perna:2016jqh}. There, one predicts an amount of matter of \numrange{.e-3}{.e-4} M$_\odot$ in a non-active accretion disk around one of the black holes, coming from a massive progenitor star with low metallicity. Upon the merger, this disk is then reactivated and leads to a burst. Using these values, one gets $\f\approx\num{.e-7}$ for the fireball model. This should be compared with the reach in Fig.~\ref{fig:fracpotentialAVUL}, rescaled to lower values of \f\ with a beaming factor. For a beaming factor of $\frac{4\pi}{0.2\times0.2}$,  this \f\ is still below the estimated reach. This is in agreement with the bound on $f_\mathrm{matter}$ found above, as the viable models predict a flux that is not yet observable. It should be noted that several other models predict that the amount of available matter would be even lower~\cite{Kimura:2016xmc}. \\

Even though neutrino production is generally not expected from BBH mergers, in this section several realistic models have been considered. It follows that the predicted neutrino fluxes are below the current limits. However, in the near future, enough BBH mergers will have been detected so that searches for neutrino emission from these sources will be able to probe the black hole binary environment, independently from searches for gamma ray emission.\\ 


Since the method presented in this article is completely general and makes no assumptions on the source properties, it can also be used for neutron star mergers and black hole-neutron star mergers. In this case one does expect an electromagnetic and neutrino emission, since there is matter present in the source environment. In fact, these objects are thought to be the inner engines of (short) GRBs~\cite{Becker:2007sv}. Note that it is also possible to have $\f>1$ for such objects. Using the decomposition of \f\ shown above, the results can be easily interpreted using specific models. 

\section{Conclusion}\label{section4}
It was investigated how the detection of GW150914 and the corresponding neutrino analysis\footnote{After the submission of this paper, the follow-up analysis for the other two GW event (candidates) was published~\cite{ANTARES:2017iky,Gando:2016zhq}. No signal neutrinos were observed. The corresponding limits can be read off from Fig.~\ref{fig:fracpotentialAVUL}.} influence the ability to constrain possible neutrino emission from BBH mergers, independent of gamma-ray observations. 
The measurements were interpreted in terms of \f, the fraction of energy released in neutrinos in a given energy range compared to the energy in gravitational waves. Additionally, under the assumption that $E_\nu$ scales linearly with $E_{\mathrm{GW}}$, the energy fraction \f\ is universal. It was shown that this assumption, in combination with a realistic BBH mass distribution flat in log mass, leads to a diffuse emission which is roughly independent of the BBH mass.\\

In our analysis isotropic emission was assumed, where the effects of beaming lead to a direct rescaling. 
The order of magnitude limits on the neutrino emission fraction \f\ are summarised in Table~\ref{tab:summary_f}.
It follows that the limits on \f\ obtained from the non-detection of counterpart neutrinos from GW150914 are weaker than those obtained from the astrophysical flux assuming a isotropic emission. In case of beamed emission, the non-detection limit could fall below the astrophysical limit.
This immediately implies that in case BBH mergers similar to GW150914 are responsible for the astrophysical neutrino flux, that either the emission from GW150914 was not beamed towards Earth, or that the beaming was smaller than $\Delta\Omega = 3.68~\mathrm{sr}$.

\begingroup
\begin{table}
\caption{\label{tab:summary_f}Summary of the strongest bounds on \f\ (order of magnitude), assuming an E$^{-2}$-power law neutrino spectrum.}
\begin{ruledtabular}
\begin{tabularx}{.15\textwidth}{XXC}
& & \f \\
\hline
\multicolumn{2}{l}{GW150914 non-detection\footnote{This bound is similar for the mono-energetic case.}} & $  \num{.e-02}\times \frac{\Delta\Omega}{4\pi}$ \\
\multicolumn{2}{l}{Astrophysical flux (GW150914-like)} & $\numrange{.e-3}{.e-2}$ \\ 
\multicolumn{2}{l}{Astrophysical flux (All LIGO mergers)} &  $\numrange{.e-3}{.e-5}$\\ 
\hline
\multicolumn{2}{l}{Prospects ($N_{\mathrm{BBH}}=10$) at 68\% CL} & $\num{.e-3}\times \frac{\Delta\Omega}{4\pi}$ \\
\multicolumn{2}{l}{Prospects ($N_{\mathrm{BBH}}=35$) at 68\% CL} & $\numrange{.e-4}{.e-3}\times \frac{\Delta\Omega}{4\pi}$ \\ 
\hline
\multicolumn{2}{l}{Expectation: Fireball + dead acc. disk\footnote{This should be compared with the bounds on \f\ for the beamed case.}} & $\num{.e-7}$
\end{tabularx}
\end{ruledtabular}
\end{table}
\endgroup

The same technique was also used to provide an estimate of the lowest \f\ that can be probed in run O2 of LIGO, by assuming all events have the same properties as GW150914. It was found that after
$N_{\mathrm{GW}}\gtrsim 10,\ 12,\ 14$ at 68\%, 95\% and 99\% CL
respectively, the \f\ that can be reached is below the highest astrophysical bound. Below this value, BBH mergers can contribute at most partially to the diffuse astrophysical neutrino flux. Estimates for the number of BBH mergers in LIGO run O2 are between 10 and 35 events. 
The average upper limits that can be reached after these numbers of events are also shown in Table~\ref{tab:summary_f}.
Furthermore, it was shown how a possible detection in the near future provides direct information about the source evolution and BBH mass distribution, as well as the neutrino emission properties. \\

The results for a more model dependent analysis were also presented. Firstly, assuming the GRB-fireball model, the current and expected bounds on \f\ were used to put a bound on the amount of matter present in the BBH environment at the time of the merger. The results of this are presented in Table~\ref{tab:summary_fmatter}. Secondly, the GRB-fireball model is combined with a model for a dead accretion disk around one of the black holes. The neutrino energy fraction expected in this situation, $\f\approx\num{.e-7}$, is below the reach of LIGO run O2. Finally, it should be noted that while for BBH mergers no neutrino emission is typically expected, realistic models of neutrino production can not be ruled out at the moment. In the future, it will be possible to use searches for neutrino emission to probe the black hole binary environment, independently from searches for gamma ray emission. In addition, the same approach can be used for other source classes, such as neutron star-black hole and neutron star-neutron star mergers, where one does expect neutrino emission. \\

\begingroup
\begin{table}[!htb]
\centering
\caption{\label{tab:summary_fmatter}Summary of the strongest bounds on $f_{\mathrm{matter}}$.}
\begin{ruledtabular}
\begin{tabularx}{.15\textwidth}{XXC}
& & $f_{\mathrm{matter}}$  \\
\hline
\multicolumn{2}{l}{GW150914 non-detection} & $\num{8e-2} \times \frac{\Delta \Omega}{0.2\times0.2}$\\
\hline
\multicolumn{2}{l}{Prospects ($N_{\mathrm{BBH}}=10$) at 68\% CL} & $\num{6e-3} \times \frac{\Delta \Omega}{0.2\times0.2}$\\
\multicolumn{2}{l}{Prospects ($N_{\mathrm{BBH}}=35$) at 68\% CL} & $\num{3e-3} \times \frac{\Delta \Omega}{0.2\times0.2}$\\
\hline
\multicolumn{2}{l}{Expectation: Fireball + dead acc. disk} & $\num{.e-7}$
\end{tabularx}
\end{ruledtabular}
\end{table}
\endgroup

\acknowledgments
We would like to thank Mauricio Bustamante, Chad Finley, Imre Bartos and Jan L\"{u}nemann for valuable discussions and feedback on the manuscript. In addition, we thank the referees for their valuable input. KDV is supported by the Flemish Foundation for Scientific Research (FWO-12L3715N - K.D. de Vries). 
GDW and JMF are supported by Belgian Science Policy (IAP VII/37) and JMF is also supported in part by IISN. MV is aspirant FWO Vlaanderen.



\bibliography{biblio}

\end{document}